\documentclass[a4paper,10pt,twocolumn]{article}
\pdfoutput=1

\usepackage[a4paper,top=2.2cm,bottom=2.4cm,left=1.8cm,right=1.8cm,columnsep=0.9cm]{geometry}
\usepackage[utf8]{inputenc}
\usepackage[T1]{fontenc}
\usepackage{microtype}      % helps with the narrow columns
\usepackage{authblk}        % same \author[a]{..} / affiliation syntax as jcappub
\usepackage{url}
\usepackage{graphicx}
\usepackage{caption}
\usepackage{subcaption}
\usepackage{amsmath}
\usepackage{amssymb}
\usepackage{multirow}
\usepackage{dirtytalk}
\usepackage{float}
\usepackage[dvipsnames]{xcolor}
\usepackage{color}
\usepackage{hyperref}
\usepackage[capitalise]{cleveref}   % must be loaded after hyperref

\usepackage{titlesec}
\titleformat{\section}{\large\bfseries}{\thesection.}{0.6em}{}
\titleformat{\subsection}{\normalsize\bfseries}{\thesubsection}{0.6em}{}
\titlespacing*{\section}{0pt}{1.4ex plus .4ex minus .2ex}{0.8ex plus .2ex}
\titlespacing*{\subsection}{0pt}{1.1ex plus .3ex minus .2ex}{0.6ex plus .2ex}

\newenvironment{acknowledgments}{\section*{Acknowledgments}}{}

\newcommand{\be}{\begin{equation}}
\newcommand{\ee}{\end{equation}}
\newcommand{\ben}{\begin{displaymath}}
\newcommand{\een}{\end{displaymath}}
\newcommand{\bea}{\begin{eqnarray}}
\newcommand{\eea}{\end{eqnarray}}

\def\be{\begin{equation}}
\def\ee{\end{equation}}
\def\bea{\begin{eqnarray}}
\def\eea{\end{eqnarray}}
\def\ba{\begin{array}}
\def\ea{\end{array}}

\def\beq{\begin{equation}}
\def\eeq{\end{equation}}

\definecolor{verde}{rgb}{0,0.5,0}

\begin{document}

% ---------------------------------------------------------------
% Title / authors / abstract: full width on top of the first page
% ---------------------------------------------------------------
\title{\bfseries\Large Tachyonic Encore:\\[2pt] A universal shift of inflationary observables}

\author[a]{Diogo S. Gorgulho}
\author[a,b]{Margherita Putti}
\author[c,a,d]{Rodrigo Gonzalez Quaglia}
\author[a]{Ema Dimastrogiovanni}
\author[e]{Matteo Fasiello}
\author[a]{Diederik Roest}

\affil[a]{Van Swinderen Institute for Particle Physics and Gravity,
University of Groningen, Nijenborgh 3, 9747 AG Groningen, The Netherlands}
\affil[b]{Dutch Institute for Emergent Phenomena, The Netherlands}
\affil[c]{Kavli Institute for the Physics and Mathematics of the Universe (WPI), The
University of Tokyo Institutes for Advanced Study, The University of Tokyo, Kashiwa,
Chiba 277-8583, Japan}
\affil[d]{Instituto de Ciencias F\'{\i}sicas, Universidad Nacional
Aut\'onoma de M\'exico, Av. Universidad s/n, Cuernavaca, Morelos, 62210, Mexico}
\affil[e]{Instituto de F\'isica T\'eorica UAM-CSIC, Calle Nicol\'as Cabrera 13-15, 28049, Madrid, Spain}
\affil[ ] {\upshape d.severino.gorgulho@rug.nl, \ m.putti@rug.nl,\ 
rodrigo.gonzalez@ipmu.jp, \ e.dimastrogiovanni@rug.nl, \
matteo.fasiello@csic.es, \ d.roest@rug.nl}
\date{}

\makeatletter
\twocolumn[
\begin{@twocolumnfalse}
\maketitle
\begin{center}
\begin{minipage}{0.88\textwidth}
\small
\noindent 
We propose a generic mechanism applicable to single-field inflaton models in which a light axion spectator, initialized near the hilltop of its potential, reshapes inflationary observables through purely gravitational multi-field dynamics. During inflation the axion is frozen and the background follows an effectively single-field trajectory. After inflation ends, the axion rolls, inducing a turn in field space and transient tachyonic phases of the isocurvature mode.~The resulting ``tachyonic encore'' occurs entirely on super-horizon scales.~These phases generate a nearly scale-invariant enhancement of the curvature power spectrum, suppressing the tensor-to-scalar ratio and shifting the scalar tilt to a weighted combination of adiabatic and entropic tilts at horizon crossing. We show that these effects can reconcile otherwise disfavored inflaton potentials with current CMB constraints.~The same dynamics predict local non-Gaussianity, $f_{\rm NL}^{\rm loc.}\sim \mathcal{O}(1)$, within reach of upcoming surveys.
\end{minipage}
\end{center}
\vspace{1.2em}
\end{@twocolumnfalse}
]
\makeatother

%%%%%%%%%%%%%%%%%%%%%%%%%%%%%%%%%%%%%%%%%%%%
%%%%%%%%%%%%%%%%%%%%%%%%%%%%%%%%%%%%%%%%%%%%

\noindent \textbf{Introduction:} Inflation, a  rapid period of accelerated expansion in the very early Universe~\cite{Linde:1990flp,Linde:2007fr,Achucarro:2022qrl,Ellis:2023wic,Kallosh:2025ijd}, stands as one of the main pillars of the standard cosmological model. It provides a mechanism for generating the primordial perturbations that seed the large-scale structure we currently observe. The simplest realizations of inflation in agreement with CMB data~\cite{Planck:2018jri,BICEP:2021xfz} are based on Einstein gravity minimally coupled to a scalar field. However, the most recent CMB observations, when combined with large-scale structure, seem to disfavor  some of the most compelling single-field scenarios~\cite{AtacamaCosmologyTelescope:2025blo}. 
We present here an intriguing multi-field mechanism that, by means of a minimally coupled axion spectator field, shifts the predicted CMB observables\footnote{Alternative approaches that have been proposed to reconcile inflationary models with the recent CMB data include shifting the horizon-crossing time \cite{Cembranos:2026cat,Kallosh:2026kfx}, considering a non-standard reheating epoch \cite{Drees:2025ngb,Ellis:2025zrf,Haque:2025uis,Zharov:2025zjg}, modifying the $N$-dependence of the observables through non-minimal couplings \cite{Kallosh:2025rni}, or modifying the inflationary potential \cite{Bezerra-Sobrinho:2025gfg,German:2026wpg}.} of the otherwise single-field configurations while also producing a primordial local non-Gaussianity signal within the reach of experiments like SPHEREx \cite{SPHEREx:2014bgr}, PUMA \cite{PUMA:2019jwd}, $21$ cm cosmology \cite{Munoz:2015eqa,Karagiannis:2019jjx} and EUCLID \cite{Yamauchi:2014ioa,Euclid:2024ris}.

In this work we will focus on multi-field scenarios\footnote{For a comprehensive review on cosmic inflation models, see e.g.~the literature in \cite{Martin:2013tda} and references therein.}, a possibility which finds strong motivation in ultraviolet finite theories such as string theory \cite{Baumann:2014nda}. Axions and axion-like particles (ALPs) in particular arise in string compactifications as zero modes of higher-form gauge fields, whose inherited shift symmetries protect their potentials and make them natural candidates for both inflation and dark matter. Their large multiplicity gives rise to the string axiverse~\cite{Arvanitaki:2009fg,Cicoli:2012sz,Demirtas:2018akl,Gendler:2023kjt}, providing a framework that connects low-energy observables to the ultraviolet structure of string theory.~ALP's possible direct couplings to the Standard Model motivate extensive experimental searches across broad masses and couplings (see~\cite{Arza:2026rsl, Albertus:2026fbe}). Naturally, such particles are also testable through their gravitational imprints, as is the case for axion inflation models~\cite{Anber:2009ua,Dimastrogiovanni:2012ew,Namba:2015gja,Peloso:2016gqs,DAmico:2021vka,DAmico:2021fhz,Dimastrogiovanni:2023juq,Pajer:2013fsa}.

In this paper we show that an axion spectator\footnote{By ``spectator'' we mean a field whose energy density is much smaller than that of the inflaton.}  can substantially reshape primordial observables even in the absence of a direct coupling with the inflaton field. During inflation, the dynamics are effectively single-field, with the axion frozen near the hilltop of its potential. After inflation ends, the axion begins to roll, inducing a turn in field space that leads to non-trivial mixing between isocurvature and curvature modes on super-horizon scales. Such dynamics includes a transient tachyonic instability, due to a negative effective isocurvature mass, which in turn leads to a net amplification of the curvature perturbation.

After COBE normalization~\cite{Planck:2018jri}, the enhanced scalar amplitude lowers the required inflationary scale and correspondingly suppresses the tensor-to-scalar ratio $r$. The curvature spectral index  is also modified with respect to the single-field  result, making it possible to reconcile otherwise disfavored models with current data. The shift in these observables is  controlled primarily by axion properties at horizon crossing, rather than by the inflaton potential. This mechanism is related to, but goes significantly beyond and generalizes
the post-inflationary enhancement  of adiabatic perturbations, discussed in the context of modular cosmology \cite{GonzalezQuaglia:2025qem}, and opens the door for novel effects on inflationary observables due to the axiverse.

On top of the CMB observables shift, the enhanced scalar spectrum we find is accompanied by large, $\mathcal{O}(1)$,  non-Gaussianity of the local type, a consequence of the sourcing of perturbations taking place solely on super-horizon scales.
Here we stress that the proposed phenomenon is very general: neither direct couplings nor non-trivial field-space geometry are necessary for the shift of CMB observables or for the generation of the local type non-Gaussianity. A systematic exploration of non-trivial field-space geometries and explicit axion-inflaton couplings is left for future work. An interesting recent work on hybrid inflation \cite{Pi:2026uxd}  also finds an enhancement of the curvature power spectrum during the post-inflationary evolution, driven by the tachyonic growth of the waterfall field and its subsequent sourcing of curvature perturbations. This enhancement can lead to significant primordial black hole production also in the context of ultra-slow roll scenarios \cite{Lorenzoni:2025kwn}.

The paper is structured as follows, in Section \ref{Multi-field inflation} we review the key aspects of the multi-field  dynamics, focusing on the sourcing of curvature fluctuations from isocurvature perturbations. Section \ref{The Axion Spectator} introduces the axion spectator sector, discusses the mass range corresponding to the \say{tachyonic encore} and describes the sourcing of the curvature perturbation from a collection of \say{tachyonic bursts} in the isocurvature sector. In Section \ref{Modified CMB Observables}, we showcase the effects of these bursts on the curvature spectrum and other CMB observables including scalar non-Gaussianity. In Section \ref{Examples} we present two concrete realizations of the mechanism for chaotic and Starobinsky inflation. We end the discussion by addressing, in Section \ref{Axion Configurations}, the naturalness of the axion configuration. Finally, Section \ref{Conclusions} contains our conclusions. In Appendix \ref{Reheating}, we show that there is a broad range of reheating temperatures for which the background remains effectively unaffected during the first few $e$-folds after the end of inflation, justifying its omission from our analysis.

%%%%%%%%%%%%%%%%%%%%%%%%%%%%%%%%%%%%%%%%%%%%%%%%%%%%%%%%%%%%%%%%%%%%%%%%%%%%%%%%%%%%%%%%%%%%%%%%%%%%%%%%%%%%%%%%%%%%%%%%

\section{Multi-field inflation}\label{Multi-field inflation} 
The generic Lagrangian for an $n$-scalar field  system has the following form 
\begin{equation}\label{MultifieldAction}
\begin{split}
    \frac{{\cal L}(\phi)}{\sqrt{-g}}={}&\frac{1}{2}R-V(\phi)-\frac{1}{2}g^{\mu\nu}G_{IJ}(\phi)\partial_{\mu}\phi^{I}\partial_{\nu}\phi^{J}\,,
\end{split}
\end{equation}
where $g$ is the determinant of the metric, $R$ is the spacetime Ricci scalar, $G_{IJ}$ is the field space metric, $\phi^I$ represents the collection of fields which in our case are the inflaton $\varphi$ and the axion $\theta$ and, in the case under scrutiny, the potential is divided as $V(\phi)=V(\varphi)+V(\theta)$.\footnote{We adopt this decomposition to provide evidence that no direct coupling between the inflaton and axion field is needed to realize the mechanism.} We have set $M_{\rm Pl}=1$.

As anticipated, the mechanism does not rely on any direct axion-inflaton coupling nor on any non-trivial field space geometry. It is merely the consequence of minimal coupling to gravity and the super-horizon evolution of perturbations characteristic of multi-field models. In the spatially-flat gauge, the quadratic Lagrangian for the perturbation $Q^I$ of the set of $n$-scalar fields reads: 
\begin{equation} \label{eomQ}
\begin{split}
    \frac{{\cal L}(Q^{I})}{a^{3} }={}&\frac{1}{2}G_{IJ}\Bigg[D_{t}Q^{I}D_{t}Q^{J}-\frac{(\partial Q^{I})\cdot(\partial Q^{J})}{a^{2}}\Bigg]-M_{IJ}^{2}Q^{I}Q^{J}\,,
\end{split}
\end{equation}
where the mass components are defined as
\begin{equation} 
\begin{split}
    M_{IJ}^{2}\equiv{}& V_{;IJ}-\mathbf{R}_{IJKL}\dot{\phi}^{K}\dot{\phi}^{L}-\frac{1}{a^{3} }D_{t}\left(\frac{a^{3}}{H}\dot{\phi}_{I}\dot{\phi}_{J}\right)\,.
\end{split}
\end{equation}
Here $\mathbf{R}_{IJKL}$ is the Riemann tensor of the field space metric constructed from $G_{IJ}$ and the overdot represents derivatives \textit{w.r.t} cosmic time. 

In order to best illustrate the analogy with the single-field  case, we define the tangential (\textit{adiabatic}) and orthogonal (\textit{entropic}) directions $\sigma$ and $s$ with respect to the background trajectory \cite{Wands:2007bd}. Specialising to the two-field case, this orthonormal basis is given by
\begin{equation}
    e_\sigma^I\equiv\frac{\dot{\phi}^I}{\dot{\sigma}} \,, \quad \mbox{and} \quad
    e_s^I \equiv - G^{-1/2} \epsilon^{IJ} \frac{\dot{\phi}_J}{\dot{\sigma}} \,,
\end{equation}
where $\dot{\sigma}^2\equiv G_{IJ}\dot{\phi}^I\dot{\phi}^J$, and  adiabatic and entropic fluctuations are given by $Q_i = e_i^I Q_I$ with $i=(\sigma,s)$. 
Importantly, these directions can evolve in time as follows:
\begin{equation}
    D_te_\sigma^I=\Omega\,e_s^I\, \quad \mbox{and} \quad  D_te_s^I=-\Omega\, e_\sigma^I\,,
\end{equation}
where we introduced the \textit{turning rate} $\Omega$ which quantifies the deviation from geodesic motion. It is convenient to define the curvature and isocurvature perturbations with normalisation as in~\cite{Bartolo_2001},
\begin{equation}\label{CurvIsoDef}
    {\cal R}\equiv \frac{1}{\sqrt{2\epsilon}}Q_\sigma\,, \quad {\cal S}\equiv\frac{1}{\sqrt{2\epsilon}}Q_s\,,
\end{equation}
where the slow roll parameters are given by
\begin{equation}
    \epsilon\equiv -\frac{\dot{H}}{H^{2}}=\frac{\dot{\sigma}^{2}}{2H^{2}}\,,\quad \eta\equiv\frac{\dot{\epsilon}}{H\epsilon}\,.
\end{equation}
While physical observables are determined by the canonically normalized variable $Q_{s}$, we will often use $\mathcal{S}$ to describe the evolution of perturbations.  The equations of motion for the set of curvature-isocurvature variables $(\mathcal{R},\mathcal{S})$ read
\begin{equation}\label{eomadi}
\begin{split}
    &\ddot{\cal R}+(3+\eta)H\dot{\cal R}+\frac{k^2}{a^2}{\cal R}=\frac{1}{a^3\epsilon}\frac{d}{dt}\left(2\epsilon a^3\frac{\Omega}{H}{\cal S}\right)\,,
\end{split}
\end{equation}
\begin{equation}\label{eomiso}
\begin{split}
    &\ddot{\cal S}+(3+\eta)H\dot{\cal S}+\left(\frac{k^2}{a^2}+m_\mathcal{S}^2\right){\cal S}=-2\frac{\Omega}{H}\dot{{\cal R}}\,.
\end{split}
\end{equation}
The mechanism presented in this paper considers an axion frozen during inflation (see examples in Section \ref{Examples}) which only rolls afterwards. Consequently, during inflation, the turn rate is negligible and no sourcing of perturbations occurs on sub-horizon scales ($k\gg aH$). This implies that the interesting phenomenology occurs entirely on super-horizon scales ($k\ll aH$), where the axion develops non-trivial dynamics such that $\Omega\neq0$. In this limit, the equations of motion for the curvature-isocurvature system simplify as follows:
\begin{equation} \label{Rdot}
\begin{split}
    &\dot{{\cal R}}=2\frac{\Omega}{H}{\cal S}+\frac{C(k)}{a^3\epsilon}\,,\\
    &\ddot{\cal S}+(3+\eta)H\dot{\cal S}+m_{\mathcal{S},{\rm eff}}^2{\cal S}={\cal O}\left(\frac{k^2}{a^2}\right)\,,
\end{split}  
\end{equation}
where $C(k)$ is a $k$-dependent integration constant, whose contribution rapidly redshifts and becomes negligible. The effective isocurvature mass reads
\begin{equation}\label{Isomass}
\begin{split}
m_{\mathcal{S},{\rm eff}}^2\equiv{}& V_{;ss}-V_{;\sigma\sigma}+\epsilon H^{2}\mathbf{R}+\frac{1}{a^3}\frac{d}{dt}\left(\frac{a^3\dot{\sigma}^2}{H}\right)+ 4\Omega^2\,,
\end{split}
\end{equation}
where $\mathbf{R}$ is the Ricci scalar of the field space metric.
In contrast to the single-field case, the curvature perturbation need not freeze on super-horizon scales; isocurvature modes can source curvature perturbations, leading to non-trivial super-horizon evolution, provided the background trajectory turns ($\Omega\not=0$).

%%%%%%%%%%%%%%%%%%%%%%%%%%%%%%%%%%%%%%%%%%%%%%%%%%%%%%%%%%%%%%%%%%%%%%%%%%%%%%%%%%%%%%%%%%%%%%%%%%%%%%%%%%%%%%%%%%%%%%%%

\section{Axion spectator and tachyonic sourcing of curvature perturbations}\label{The Axion Spectator}

The essential requirements for the encore mechanism are remarkably minimal. First, effective single-field dynamics during inflation ($\Omega\simeq 0$) with an axion spectator field frozen in a region of negative potential curvature. A simple example is the axion potential
\begin{equation}
    V(\theta)=\Lambda^4\left[1- \cos\left(\frac{\theta}{f}\right)\right]\,,
\end{equation}
where $f$ is the axion decay constant and $\Lambda$ the scale of the potential. Second,   an axion mass range
satisfying\footnote{The \say{encore} here takes place before reheating becomes efficient. This is possible for a broad range of viable reheating temperatures \cite{Gorbunov:2010bn,Dai:2014jja,Cook:2015vqa,Eshaghi:2016kne,German:2022sjd,Garcia:2023tkk,German:2023yer,German:2025mzg,Ellis:2025zrf}. See also \textit{Appendix} \ref{Reheating}.}
\begin{equation}\label{CondMass}
    H_{\rm reh}\lesssim |m_{\mathcal{\theta}}| \simeq H_{\rm end}\ll H_{\rm hc}\,,
\end{equation}
where $m_\theta^2 \equiv V_{;\theta \theta}|_{\rm hilltop} = - \Lambda^4/f^2 < 0$ and the quantities $H_{\rm reh}$, $H_{\rm end}$ and $H_{\rm hc}$ denote respectively the Hubble scale during reheating, at the end of inflation and at horizon crossing of the CMB modes. This hierarchy ensures that the axion remains frozen  during inflation ($|m_\theta| \ll H_{\rm hc}$), begins to roll shortly after its end ($|m_\theta| \simeq H_{\rm end}$), and does so before reheating becomes efficient ($|m_\theta| \gtrsim H_{\rm reh}$), thereby allowing the enhancement to occur entirely on super-horizon scales. These bounds can be directly expressed in terms of more fundamental parameters via the relation $|m_\theta|\sim\Lambda^2/f$.

%%%%%%%%%%%%%%%%%%%%%%%%%%%%%%%%%%%%%%%%%%%%%%%%%%%%%%%%%%%%%%%%%%%%%%%%%%%%%%%%%%%%%%%%%%%%%%%%%%%%%%%%%%%%%%%%%%%%%%%%

Once inflation ends, the axion field starts rolling towards its minimum, with the isocurvature mass squared repeatedly transitioning between negative and positive values
(for concrete examples see \cref{fig:BackgroundMass}). Correspondingly, isocurvature perturbations  undergo transient growth followed by a reduction in amplitude. At the same time, the field space trajectory is non-geodesic, with a non-zero turn-rate.~As dictated by \cref{Rdot}, each burst in isocurvature fluctuations is transferred to the scalar curvature.~This process continues until the isocurvature mode is sufficiently damped or the axion exits the region of negative curvature. The cumulative effect of these bursts is a net enhancement of the curvature perturbation, which eventually settles into a new plateau (see \cref{fig:PowerEnhancement}). 

The post-enhancement amplitude of the power spectrum is related to its value at the end of inflation via the \textit{enhancement parameter},
 \begin{equation}\label{defenhancement}
    {\cal E}\equiv\frac{{\cal A}^{\rm s}_{{\cal R}}}{{\cal A}^{0}_{{\cal R}}}\; ,
 \end{equation}
where $\mathcal{A}_{\mathcal{R}}^{0}$  and $\mathcal{A}_{\mathcal{R}}^{\rm s}$ are the amplitudes of the scalar power spectrum at the end of inflation and after the enhancement ends, respectively.

One can arrive at an expression for the enhancement via the $\delta N$ formalism~\cite{Sasaki:1995aw,Lyth:2004gb}. The key variable is the fluctuation in the number of $e$-folds between an initial spatially flat slice and a final uniform density one which, on super-horizon scales, corresponds to the curvature perturbation $\mathcal{R}=N(\varphi,\theta)-\bar{N}$. Expanding to first order, we find
\begin{equation}
    \mathcal{R}_{1}=N_{\varphi}\delta\varphi+N_{\theta}\delta\theta\, , 
\end{equation}
where $N_{X}=\partial N/\partial X$. As the inflationary dynamics are effectively single-field, the fluctuations are uncorrelated at horizon crossing, yielding 
\begin{equation}\label{FinalAs}
    \mathcal{A}_{\mathcal{R}}^{s}=\left(\frac{H_{\rm hc}}{2\pi}\right)^{2} \left(N_{\varphi}^{2}+N_{\theta}^{2}\right) \,, 
\end{equation}
and hence the enhancement parameter becomes
\begin{equation}\label{EnhancementDef}
    \mathcal{E} \equiv \frac{A_{\mathcal R}^{s}}{A_{\mathcal R}^{0}}
    = 1 + \frac{N_\theta^2}{N_\varphi^2}\, .
\end{equation}
This quantity controls the shift of the inflationary observables induced by the post-inflationary evolution. In general, the full dependence of $N(\varphi,\theta)$ on the post-inflationary dynamics must be determined numerically. Nevertheless, the parametric dependence of the enhancement can be understood analytically by approximating the axion as an overdamped oscillator.

The inflaton contribution $N_\varphi$ is fixed by the inflationary background; for instance, $N_\varphi=\varphi/2$ for quadratic inflation. The dependence on the axion initial condition can instead be approximated as
\begin{equation}\label{Enne}
\begin{split}
    N(\Delta\theta)= -2(1+\omega) f^2
    \log\left[
        \sin\left(\frac{\Delta\theta}{2f}\right)
    \right]\, ,
\end{split}
\end{equation}
where $\Delta\theta\equiv \pi f-\theta_{\rm in}$ measures the initial displacement from the hilltop. For $\omega=0$, this expression reduces to the overdamped single-axion result. We use $\omega \simeq V(\varphi)/V(\theta)$ during the encore to account approximately for the additional Hubble friction due to the inflaton sector.

Treating $\omega$ as approximately constant during this stage, \cref{Enne} gives
\[
    N_\theta
    = \frac{\partial N}{\partial \theta_{\rm in}}
    = (1+\omega)f
    \cot\left(\frac{\Delta\theta}{2f}\right).
\]
Close to the hilltop, $\Delta\theta/f\ll 1$, and therefore
\[
    N_\theta \simeq
    \frac{2(1+\omega)f^2}{\Delta\theta}\, .
\]
Substituting this result into \cref{EnhancementDef}, and focusing on the regime in which the axion contribution dominates, $\mathcal{E}\gg1$, we obtain
\begin{equation}\label{EnhancementintermsofN}
    \mathcal{E} \simeq
    \frac{4f^4(1+\omega)^2}
         {N_\varphi^2\,\Delta\theta^2}\, .
\end{equation}
The enhancement therefore grows as $\Delta\theta^{-2}$: initial conditions closer to the hilltop lead to a larger amplification of the curvature perturbation.

%%%%%%%%%%%%%%%%%%%%%%%%%%%%%%%%%%%%%%%%%%%%%%%%%%%%%%%%%%%%%%%%%%%%%%%%%%%%%%%%%%%%%%%%%%%%%%%%%%%%%%%%%%%%%%%%%%%%%%%%

\section{Modified CMB observables and the generation of non-Gaussianities }\label{Modified CMB Observables} 
\subsection{Scalar amplitude and tensor-to-scalar ratio}
We next examine one of the two main effects of the mechanism presented in this paper: shifting the CMB observables without modifying the inflaton potential. A direct consequence of the enhancement of the scalar power spectrum is that the COBE normalization, $\sim 2.1\times10^{-9}$, must be matched by the post-enhancement amplitude $\mathcal{A}_{\mathcal R}^{s}$ (see Fig. \ref{fig:PowerEnhancement}). From the definition $\mathcal{E} \equiv \mathcal{A}_{\mathcal R}^{s}/\mathcal{A}_{\mathcal R}^{0}$, the amplitude $\mathcal{A}_{\mathcal R}^{0}$ at the end of inflation is therefore smaller than the final amplitude by a factor $\mathcal E$. Since the spectator axion does not affect the tensor spectrum, the enhancement of the scalar sector directly suppresses the tensor-to-scalar ratio relative to its single-field value, with the suppression controlled by $\mathcal E \geq 1$, 
\begin{equation}\label{Finalr}
    r^{\rm post}=\frac{\mathcal{P}_{T}}{\mathcal{P}^{s }_{\mathcal{R}}}=\frac{\mathcal{P}_{T}}{\mathcal{E}\  \mathcal{P}_{\mathcal{R}}^{0}}\,.
\end{equation}
The degree of suppression of the tensor-to-scalar ratio is therefore set by the efficiency of the post-inflationary transfer from isocurvature to curvature perturbations.

\subsection{Scalar spectral index}

As we previously discussed, the inflationary dynamics are single-field like. Therefore, if we were to evaluate the scalar spectral index at the end of inflation (or at horizon crossing) we would find the single-field result 
\begin{equation}
    n_{\mathcal{R}_{0}}-1 = 2\eta_{\sigma\sigma} -6\epsilon \,,
\end{equation}
where $\eta_{\alpha\alpha}\equiv V_{;\alpha\alpha}/V$. This expression does not account for the intrinsic scale dependence of the axion (isocurvature) sector, which contributes to the final curvature spectrum once the encore becomes active. Since the transfer occurs entirely on super-horizon scales, it does not introduce additional $k$-dependence. 

During inflation, both the curvature and isocurvature modes acquire their intrinsic scale dependence at horizon crossing. Their contributions are then mixed during the encore through the super-horizon sourcing of curvature perturbations. The final scalar tilt is therefore determined by the relative contributions of the curvature and isocurvature spectra evaluated at horizon crossing,
\begin{equation}\label{Finalns}
    n_{s}^{\rm post} = \frac{1}{\mathcal{E}}n_{\mathcal{R}_{0}}  + \frac{\mathcal{E}-1}{\mathcal{E}} n_{Q_s} \Bigg|_{\rm hc}\,.
\end{equation}
Equivalently, this expression can be obtained directly from the post-enhancement amplitude in Eq.~\eqref{FinalAs} by taking its logarithmic derivative with respect to $k$. The enhancement factor determines whether the inflaton or axion sectors dominate the spectral tilt while the special case $\mathcal{E}=1$ (no enhancement) reproduces the single field result. 

The entropic mode (and its equation of motion) corresponds to a massive field in a quasi-de Sitter space (see \textit{e.g}.~\cite{Riotto:2002yw}) so that its power spectrum is given by
\begin{equation}
    \mathcal{P}_{Q_{s}}=H_{\rm hc}^{2}\left(\frac{k}{k_{*}}\right)^{\nu} \,, \quad \nu \simeq 2 \left( \eta_{s s}+ \frac{\epsilon \mathbf{R}}{3} \right) \,,
\end{equation}
with spectral index 
\begin{equation}\label{NewSpectralIndex}
\begin{split}
     n_{Q_{s}}-1&=\frac{d\log \mathcal{P}_{Q_{s}}}{d\log k}= 2\eta_{ss} + \left(\frac{2}{3}\mathbf{R}-2\right)\epsilon\,.
\end{split}
\end{equation} 
The field-space geometry enters the isocurvature tilt explicitly through the Ricci scalar $\mathbf{R}$ and can therefore affect its precise value. In the flat field-space realizations considered below, \(\mathbf R=0\), so this contribution vanishes. More generally, the mechanism does not rely on a specific choice of geometry; for curved field spaces, the geometric correction is proportional to $\epsilon\mathbf R$ and is therefore subleading, provided the field-space curvature is not parametrically large.

Beyond the modification of the spectral index, the weak $k$-dependence arising from the evaluation of the background quantities at horizon crossing can also affect the running of the spectral index. Differentiating Eq. (4.3) with respect to $\ln k$, we obtain
\begin{equation*}
\begin{split}
    \alpha^{\rm post}_{s}
    ={}&\frac{\alpha_s}{\mathcal{E}}
    +\frac{\mathcal{E}-1}{\mathcal{E}}\alpha_{Q_s}-\frac{n_{\mathcal{R}_0}-n_{Q_s}}{\mathcal{E}^{2}}
    \frac{d\mathcal{E}}{d\ln k}\,.
\end{split}
\end{equation*}
In the large-enhancement regime, the running is dominated by the isocurvature contribution, $\alpha_s^{\rm post}\simeq\alpha_{Q_s}$, up to subleading corrections. A detailed analysis of the running and its observational implications is left for future work.

\cref{Finalr,Finalns} depict the generic form of the modified CMB observables for a single-field inflaton potential in the presence of an axion spectator. Whenever the axion dynamics dominate, $\mathcal{E}\gg1$, these observables generically deviate from their single-field predictions, potentially opening up regions of parameter space in which otherwise disfavored single-field potentials become compatible with recent data.

Interestingly, Ref.~\cite{GonzalezQuaglia:2025qem} showed that, for a two-field model with a hyperbolic field-space metric and an $\alpha$-attractor-like inflaton potential, the post-enhancement spectral index agrees at leading order with the standard single-field $\alpha$-attractor prediction. There, this result was obtained from the $N$-scaling of $\mathcal{A}_{\mathcal{R}}^{0}$  and $\mathcal{A}_{\mathcal{R}}^{s}$, which both scale as $N^2$ and therefore yield the same leading $1/|N|$ contribution to the spectral index. Here, we recover this result directly from the general expression in \cref{Finalns}. We first calculate the entropic tilt $n_{{Q}_{s}}$, considering a hyperbolic field-space metric given by 
\begin{equation}
    G_{IJ}={\rm diag} \left( 1\,,\,\frac{3\alpha}{2}\,\rm exp\left[-2\sqrt{\frac{2}{3\alpha}}\,\varphi\right]\right)\,,
\end{equation}
where $\alpha$ is a free parameter related to the curvature of the K\"{a}hler manifold. Since the entropic direction is aligned with the axion direction during inflation, we compute the leading contribution to $\eta_{ss}\simeq \eta_{\theta\theta}=-1/|N|$, using the known $\alpha$-attractor relation $\epsilon=3\alpha/(4N^2)$. 
Following \cref{NewSpectralIndex} and accounting for the contribution of the non-trivial field-space metric, we find
\begin{equation}
n_{Q_{s}}=1-\frac{2}{|N|}+\mathcal{O}(1/N^2)\,.
\end{equation}
Finally, recalling the universal $\alpha$-attractor result, $n_{\mathcal{R}_{0}}\sim 
1-2/|N|$, we notice that the adiabatic and entropic modes have the same leading order tilt. Therefore, regardless of how they mix and independently of the value of the enhancement factor, the observed spectral index $n_{s}^{\rm post}$ retains the same, leading order form, thus recovering the result in \cite{GonzalezQuaglia:2025qem},
\begin{equation}
n_{s}^{\rm post}\simeq n_{\mathcal{R}_{0}}\simeq 
1-\frac{2}{|N|}\,.
\end{equation}
It would be interesting to see whether this leading-order matching between the adiabatic and entropic tilts persists beyond this example.

\subsection{Generation of non-Gaussianity}
Non-Gaussianity is often generated by interactions that make the field fluctuations intrinsically non-Gaussian. Here, instead, we neglect the small intrinsic non-Gaussianity present at horizon crossing and focus on the dominant contribution generated by the subsequent multi-field super-horizon evolution. This contribution arises from the non-linear dependence of the number of $e$-folds on the field values (see \cref{Enne}).

Recalling that the enhancement dynamics takes place entirely on super-horizon scales, we anticipate (and also verify numerically) that this mechanism engenders\footnote{This contribution is essentially additive with respect to any other non-Gaussianity stemming from inflationary (self-)interactions. It is well-know that derivative interactions support equilateral and orthogonal type non-Gaussianity whilst standard single-field slow-roll dynamics driven by a simple $V(\varphi)$ leads to a slow roll suppressed local bispectrum \cite{Maldacena:2002vr}.} a local component for the bispectrum which is well-captured by the $\delta N$ formalism. The second order expansion of the curvature perturbation  reads  
\begin{equation}
\begin{split}
    \mathcal{R}_{2}={}&\frac{1}{2}N_{\varphi\varphi}\delta\varphi^{2}+\frac{1}{2}N_{\theta\theta}\delta\theta^{2}+N_{\varphi\theta}\delta\varphi\delta\theta\,.
\end{split}
\end{equation}
Computing the bispectrum as $\langle \mathcal{R}_{1}\mathcal{R}_{1}\mathcal{R}_{2} \rangle + \text{2 perm.}$\footnote{Notice that $\langle\mathcal{R}_{1}\mathcal{R}_{1}\mathcal{R}_{1}\rangle=0$ as $\mathcal{R}_1$ is linear in Gaussian fields fluctuations.} and comparing with the standard local non-Gaussian definition \cite{Lyth:2005fi}, we find the following form of the non-linear parameter $f_{\rm NL}$.\footnote{This is the dominant super-horizon contribution for gravitationally interacting fields, as discussed in \cite{Iarygina:2023msy}.} 
\begin{align}\label{Finalfnl}
f^{\rm loc.}_{\mathrm{NL}} &= \frac{5}{6}\,\Bigg[\nonumber\frac{ N_{\varphi\varphi} N_\varphi^2 + N_{\theta\theta} N_\theta^2+ 2 N_{\varphi\theta} N_\varphi N_\theta  }{ \left(N_\varphi^2 + N_\theta^2\right)^2 }\Bigg] \nonumber\\[4pt]
&= \frac{5}{6} \Bigg[ \frac{1}{\mathcal{E} ^2}\frac{N_{\varphi\varphi}}{N_\varphi^2} \nonumber+ \frac{2(\mathcal{E}-1)}{\mathcal{E}^2}\frac{N_{\varphi\theta}}{N_\varphi N_\theta} \nonumber+ \left(\frac{\mathcal{E}-1}{\mathcal{E}}\right)^2 \frac{N_{\theta\theta}}{N_\theta^2} \Bigg] \; .
\end{align}
This formula, analogous to \cref{Finalr,Finalns}, is modulated by the enhancement parameter. The first term in this expression depends only on the inflaton sector and is slow-roll suppressed, $\mathcal{O}(\epsilon,\eta)$ \cite{Maldacena:2002vr}, with an additional $1/\mathcal E^2$ suppression. The second term scales as $\mathcal O(1/\mathcal E)$ for $\mathcal E\gg1$. The third term is controlled entirely by the axion dynamics and, using \cref{Enne}, gives an $\mathcal O(1)$ contribution in this regime,
\begin{equation}\label{FinalFnl}
    f_{\rm NL}^{\rm loc.}\simeq\frac{5}{6}\frac{N_{\theta\theta}}{N_{\theta}^{2}}\simeq\frac{5}{12}\frac{1}{(1+\omega) f^2}\;.
\end{equation}
In contrast to the curvaton \cite{Bartolo:2002vf,Byrnes:2025kit} or modulated reheating \cite{Dvali:2003em,Dvali:2003ar}, the generation of $\mathcal{O}(1)$ local non-Gaussianity in our scenario does not require an additional conversion mechanism; both the enhancement and its non-linear signature arise from the same post-inflationary dynamics.

Given current CMB bounds on $f_{\rm NL}$ \cite{Planck:2018nkj,Planck:2019kim}, our result is rather consequential on the axion decay constant allowed range.\footnote{In string-theory constructions $f$ is typically sub-Planckian and not far below $M_{\rm Pl}$,  often lying within a few orders of $M_{\rm Pl}$.}

\section{Chaotic and Starobinsky realizations}\label{Examples}

In this section we present two specific realizations of the tachyonic encore mechanism, employing the chaotic \cite{Linde:1983gd} and Starobinsky \cite{Starobinsky:1980te} potentials in a flat field-space geometry. We stress that the mechanism is not tied to a specific inflaton potential or field-space geometry, and that the examples considered here provide two minimal realizations based on representative, well-known potentials.

\begin{figure*}[t]
\centering
\begin{subfigure}{0.5\linewidth}
\centering
\includegraphics[width=1\textwidth]{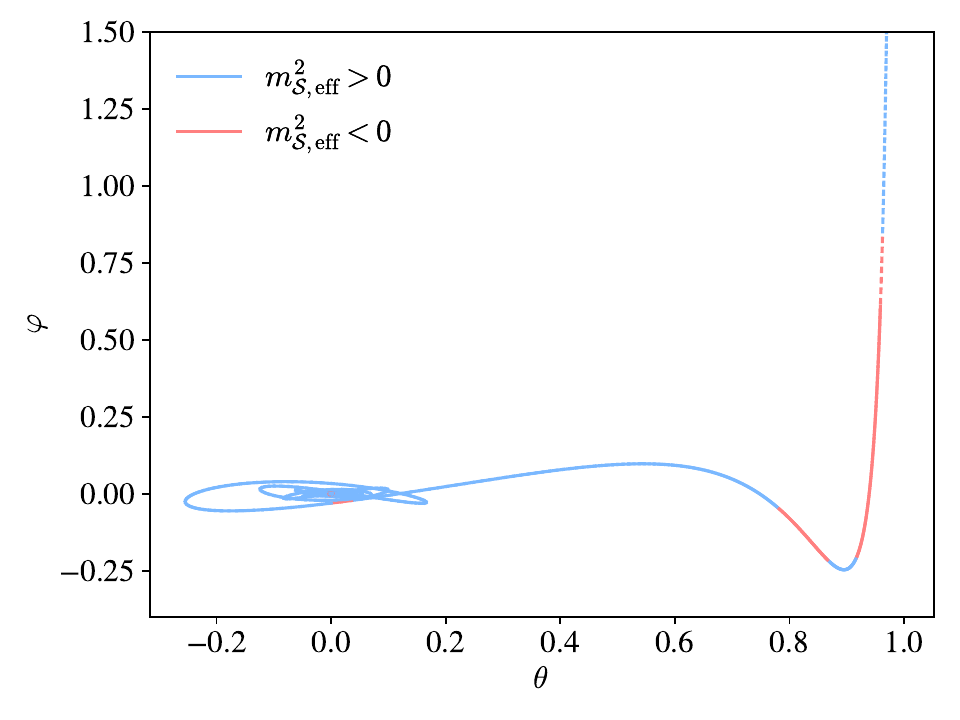}
\end{subfigure}%
\begin{subfigure}{0.5\linewidth}
\centering
\includegraphics[width=1\textwidth]{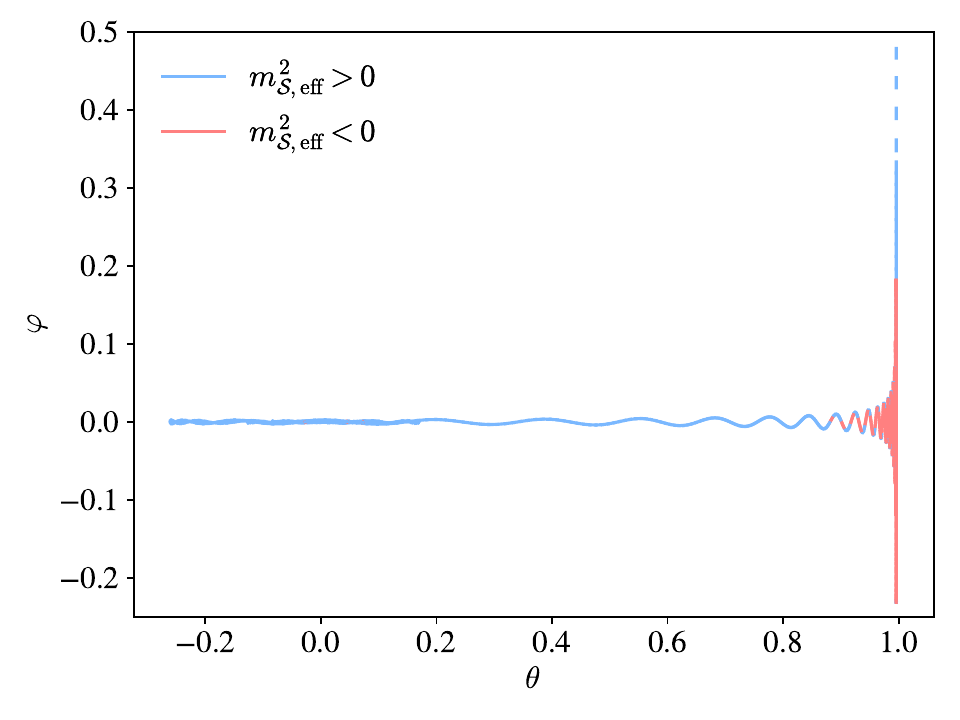}
\end{subfigure}
\caption{Field-space trajectory for chaotic (left) and Starobinsky (right) inflation potentials with parameters given in \cref{parameters_chaotic} and \cref{parameters_Staro}, respectively. The (solid) dashed  line corresponds to (post-)inflation with ($N>0$) $N<0$. Note that inflation is effectively single-field and the effective isocurvature mass alternates between stable and tachyonic regimes afterwards.}
\label{fig:BackgroundMass}
\end{figure*}

\begin{figure*}[t!]
\centering
\begin{subfigure}{0.5\linewidth}
\centering
        \includegraphics[width=1\textwidth]{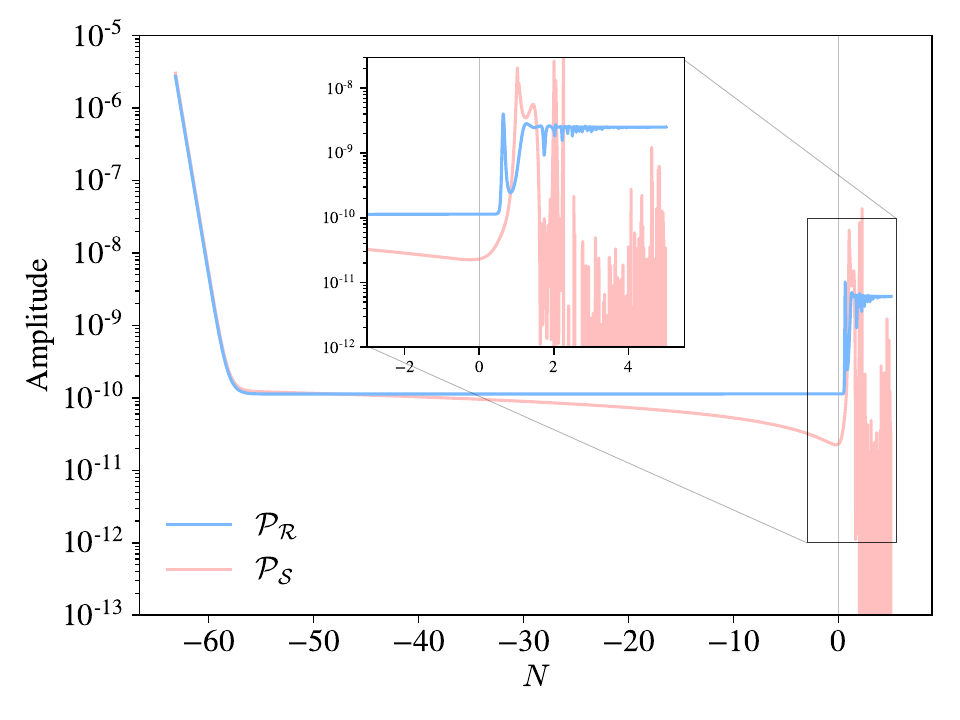}
\end{subfigure}%
\begin{subfigure}{0.5\linewidth}
\centering
        \includegraphics[width=0.99\textwidth]{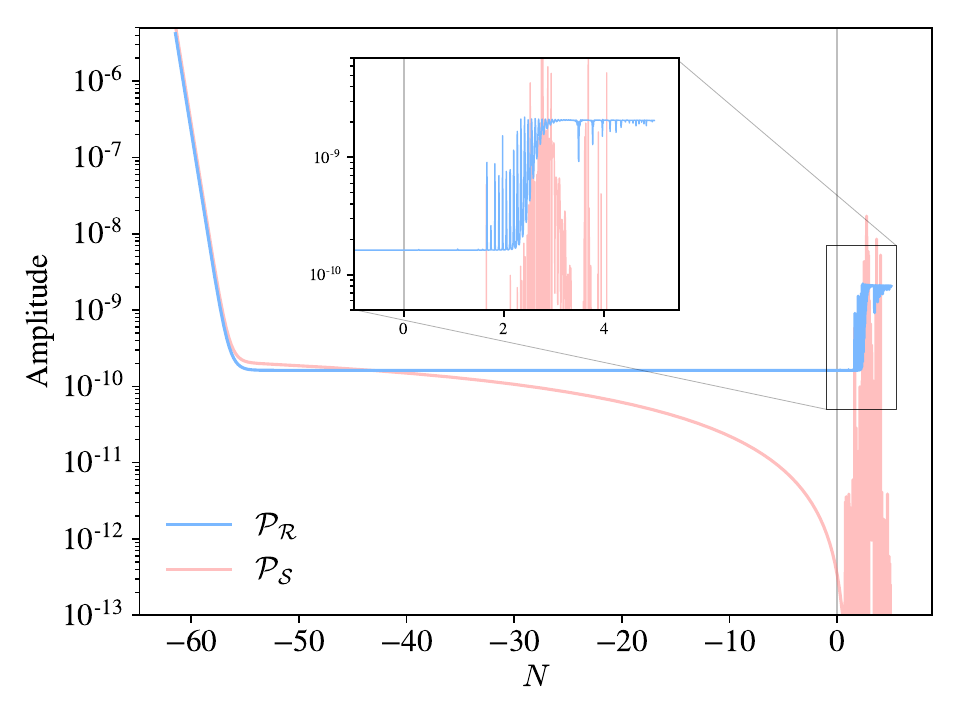}
\end{subfigure}
    \caption{Amplitude of the power of curvature and isocurvature fluctuations of the CMB $k$-mode ($N_{\rm hc} \simeq -60$) for chaotic (left) and Starobinsky (right) inflation with parameters given in \cref{parameters_chaotic} and \cref{parameters_Staro}, respectively, illustrating the bursts and decays of isocurvature fluctuations and the sourcing of curvature fluctuations. Plots obtained using \texttt{PyTransport}~\cite{Mulryne:2016mzv}.
    }
    \label{fig:PowerEnhancement}
\end{figure*}

\begin{figure*}[t]
    \centering
    \includegraphics[trim=55pt 20pt 80pt 20pt, clip,width=0.95\textwidth]{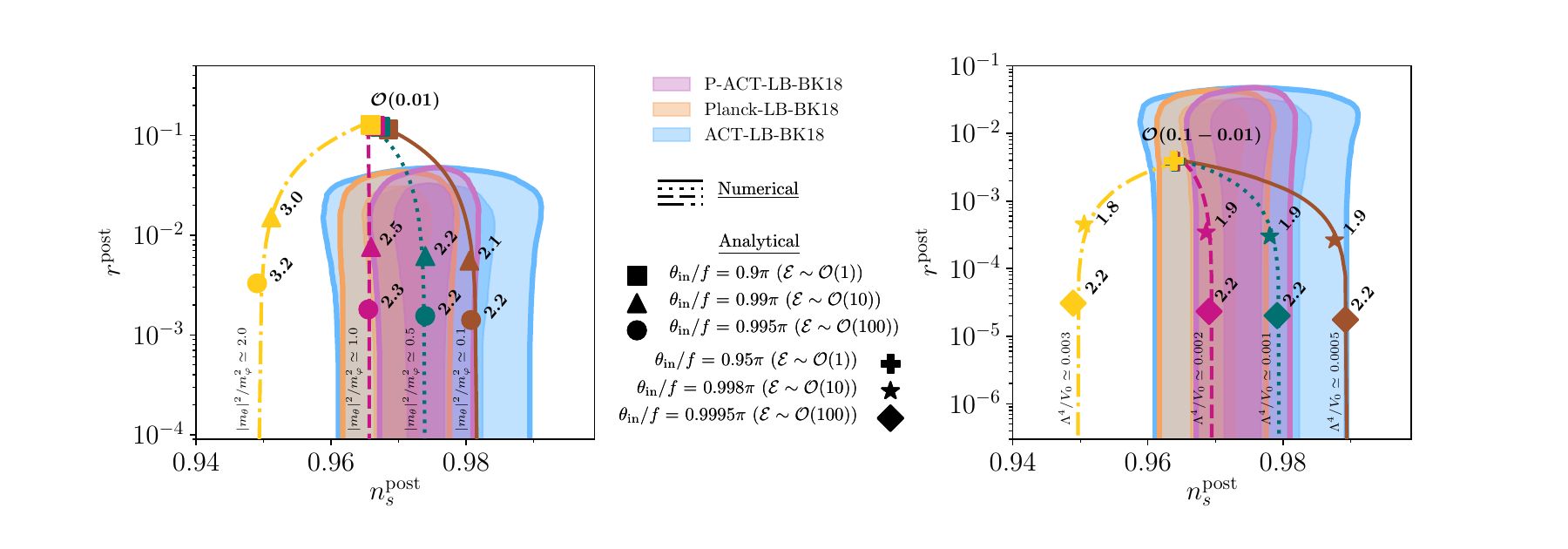} 
    \caption{$n_{s}-r$ plane (for $f=1/\pi$) with the numerical (lines) and analytical (markers) predictions for chaotic (left) and Starobinsky (right) inflation with flat internal geometry, together with the $2\sigma$ confidence regions of diverse data sets. We include, next to each prediction set, the values of the corresponding $f_{\rm NL}^{\rm loc.}$ (in boldface), calculated numerically with \texttt{PyTransport}, which sit inside the latest bounds \cite{Planck:2018nkj,Planck:2019kim,Chaussidon:2024qni,Rosado-Marin:2026kpx}. Notice that the encore is effective whenever $\mathcal{E}\gg1$. In this regime, $n_{s}^{\rm post}$ is shifted from the standard single-field results by corrections controlled by $m_{\theta}^2/m_{\varphi}^2$ (for chaotic) and $\Lambda^4/V_0$ (for Starobinsky), $r^{\rm post}$ is suppressed, and $f_{\rm NL}^{\rm loc.}\sim \mathcal{O}(1)$.
   In the complementary regime $\mathcal{E}\approx1$ we recover the standard inflation observables, which in the chaotic case lie outside the confidence regions of the diverse data sets, and an $\mathcal{O}(\epsilon,\eta)$ non-Gaussian signal.}
    \label{fig:ns_r}
\end{figure*}

The Lagrangian for these examples reads
\begin{equation}
\begin{split}
    \frac{{\cal L}(\varphi,\theta)}{\sqrt{-g}}={}&\frac{1}{2}R-V(\varphi)-V(\theta)-\frac{1}{2}\partial_{\mu}\varphi\partial^{\mu}\varphi-\frac{1}{2}\partial_{\mu}\theta\partial^{\mu}\theta\,.
\end{split}
\end{equation}
In Fig.~\ref{fig:BackgroundMass} we show the field-space motion of both the inflaton and axion fields for the following sets of fiducial parameters\footnote{These numerical values do not require a special tuning of the spectator dynamics: as long as the axion mass satisfies the hierarchy in \cref{CondMass}, the axion remains frozen during inflation. The specific parameter choices adopted here are chosen to yield predictions compatible with current observations (see Fig.~\ref{fig:ns_r}).}:

\noindent 
\textbf{Chaotic potential:}
\begin{align}\label{parameters_chaotic}
   & V(\varphi) = \frac{1}{2} m_\varphi^2 \varphi^2 \,, \nonumber \\
   & \theta_{\rm in} / f=0.99\pi \,, \quad \Lambda^4/m_\varphi^2=0.05 \,,\quad f=1/\pi \,.
 \end{align}
\textbf{Starobinsky potential:}
\begin{align}\label{parameters_Staro}
   & V(\varphi) = V_0 \left( 1 - e^{-\sqrt{2/3}\,\varphi} \right)^2 \,, \nonumber \\
   & \theta_{\rm in} / f= 0.998\pi\,, \quad \Lambda^4/V_0= 0.001 \quad f=1/\pi \,. 
 \end{align}
In both cases, the axion remains frozen near the hilltop of its potential until the end of inflation, so that the inflationary dynamics are effectively single-field and the sourcing of curvature perturbations is negligible. After inflation, the isocurvature mass squared changes sign, giving rise to the transient tachyonic phases responsible for the enhancement of the curvature power spectrum shown in Fig.~\ref{fig:PowerEnhancement}.

The phenomenologically interesting regime is the one in which the CMB observables deviate from their single-field predictions, namely when the axion contribution dominates in \cref{FinalAs,Finalr,Finalns}. In the limit $\mathcal{E}\gg 1$, these expressions reduce to the following analytic forms:

\noindent\textbf{Chaotic potential:}
\begin{equation}\label{Observables1}
\begin{split}
 \mathcal{A}_s^{\rm post}
    &\simeq\frac{8f^4}{3\pi^2\Delta\theta^2}m_\varphi^2|N|\,,\\
    r^{\rm post}&\simeq\frac{\Delta\theta^2}{2f^4}\,, \quad
    n_s^{\rm post}
    \simeq
    1-\frac{1}{|N|}\left(1-\frac{m_{\theta}^2}{m_\varphi^2}\right)\,.
\end{split}
\end{equation}
\textbf{Starobinsky potential:}
   \begin{equation}\label{Observables2}
\begin{split}
\mathcal{A}_s^{\rm post}
    &\simeq\frac{4f^{4} V_{0}}{3\pi^2 \Delta\theta^{2} }\left(1-\frac{3}{2|N|}\right)\,,\\
    r^{\rm post} &\simeq\frac{\Delta\theta^2}{2f^4}\,, \quad
    n_s^{\rm post}\simeq
    1-\frac{\Lambda^{4}}{f^{2}V_{0}} \left(2 + \frac{3}{|N|}\right)-\frac{3}{2N^{2}}\,.
\end{split}
\end{equation}
In Fig.~\ref{fig:ns_r} we show the numerical and analytical predictions for $n_{s}$ and $r$ together with the non-Gaussian parameter in \cref{FinalFnl}, where we have also fixed $\mathcal{A}_{s}^{\rm post}$ to the COBE normalization. 

As already anticipated, in the large enhancement regime, the CMB data primarily constrain the axion sector rather than the inflaton one. In particular, the mass and $(\Lambda^{4}/V_{0})$ ratio control the shift in $n_s$, making it possible to move towards larger values, $\Delta\theta$ determines $r$ while the COBE normalization sets the inflaton mass and the potential scale correspondingly.

Finally, although a comprehensive parameter scan or statistical analysis is beyond the scope of this paper, we provide a first analytical estimate of the region of parameter space compatible with current observations. To this end, we invert the expressions for the observables given above, including \cref{FinalFnl}, evaluate them at \(|N|=60\) and impose the constraints from the latest data sets \cite{Planck:2018jri,BICEP:2021xfz,AtacamaCosmologyTelescope:2025blo,ACT:2025tim}:
\begin{align}
    &\textbf{Chaotic potential:}\nonumber\\
    &\quad \Delta\theta\lesssim10^{-3}\,,\quad f\gtrsim\frac{0.44}{\pi}\,,\quad 0.13\lesssim
    \frac{|m_\theta|^2}{m_\varphi^2}
    \lesssim 0.95 \,.\\
    &\textbf{Starobinsky potential:}\nonumber\\
    &\quad \Delta\theta\lesssim10^{-2}\,,\quad f\gtrsim\frac{0.44}{\pi}\,,\quad 5\times 10^{-4}\lesssim
    \frac{\Lambda^{4}}{V_{0}}
    \lesssim 2\times 10^{-3}\,. 
\end{align}
These ranges therefore provide a simple motivation for the benchmark parameters adopted in \cref{fig:ns_r}. They should not be interpreted as statistical constraints, but rather as an indication that the relevant phenomenology can be realized without a precise tuning of the model parameters.

%%%%%%%%%%%%%%%%%%%%%%%%%%%%%%%%%%%%%%%%%%%%%%%%%%%%%%%%%%%%%%%%%%%%%%%%%%%%%%%%%%%%%%%%%%%%%%%%%%%%%%%%%%%%%%%%%%%%%%%%

\section{Axion configurations}\label{Axion Configurations}
So far, of the two central inputs of the mechanism, the near hilltop initial condition and the mass hierarchy in \cref{CondMass}, the former has been imposed to initialize the numerical evolution, while the latter is satisfied by the fiducial set of parameters discussed in the previous section. Although our main goal is to illustrate the mechanism rather than justify a first-principles origin of these inputs, we provide evidence that this initial condition and mass hierarchy can arise in plausible configurations.

\subsection{Stochastic framework}
Let us first address the hilltop initial condition of the axion. Considering the axion to be effectively massless throughout inflation, we can assume that its initial value is uniformly distributed over its periodic field range $0<\theta<2\pi f$. In such case, the probability of initially finding the field within a distance $\Delta\theta$ of the hilltop scales linearly with $\Delta\theta$.\footnote{We thank Andrei Linde for enlightening discussions on this topic.}  Therefore, as the encore relies on small displacements, we can conclude that initial configurations near the hilltop are linearly suppressed. For example, configurations satisfying $\Delta\theta\sim10^{-2}$ occur with probability $\mathcal{O}(10^{-2})$.

One may worry that stochastic fluctuations during inflation could significantly alter this estimate \cite{Nambu:1987ef,Nambu:1988je,Mollerach:1990zf}.
 Their effect can be incorporated through a stochastic noise term describing the quantum kicks to the classical field evolution. The axion dynamics are then governed by a Langevin equation \cite{Starobinsky:1986fx,Starobinsky:1994bd,Vennin:2015hra}
    \begin{equation}
    \frac{d\theta}{dt}=-\frac{V_\theta} {3H}+\xi(t),
\end{equation}
where $\xi(t)$ is a stochastic noise term.
Close to the hilltop, the contribution from the potential is subdominant and the axion dynamics is governed by the diffusion process, giving rise to a Gaussian probability distribution function (PDF) for the axion. This distribution, $P(\Delta\theta,N_{\rm sto})$, can
be interpreted as the probability of realizing a given axion
displacement $\Delta\theta$ after $N_{\rm sto}$ $e$-folds of stochastic evolution.
However, the stochastic diffusion time required to populate the full periodic field range and approach a flat equilibrium distribution is much longer than the duration of inflation, $N_{\rm eq}\propto (2\pi f/H)^2\sim \mathcal{O}(10^8)$ \cite{Hardwick:2017fjo}. Consequently, over the $\mathcal{O}(10^2)$ $e$-folds relevant for inflation, the axion remains localized around its initial value and only explores a small neighborhood in field space. The probability of ending inflation sufficiently close to the hilltop, and thus realizing the encore, remains therefore controlled by the initial measure factor discussed above, up to small diffusion-induced corrections.
Indeed, in the diffusion regime, the axion is governed by a Gaussian PDF with variance $\sigma_{\theta}^{2}=H_{\rm hc}^{2}N_{\rm sto}/4\pi\sim 10^{-8}$. The small variance therefore ensures that the PDF remains narrowly peaked around the initial axion value, so that stochastic diffusion does not significantly displace the field from the hilltop and the initial conditions required for the encore are preserved.

\subsection{Axiverse}
We find it useful to look at the tachyonic encore mechanism also within the context of the string axiverse~\cite{Arvanitaki:2009fg,Cicoli:2012sz,Demirtas:2018akl,Gendler:2023kjt}, where the axion at the heart of the encore is part of a set of additional
spectator sectors.
The presence of several such
axions  makes the required hilltop initial condition and mass range  substantially less restrictive than in a one-spectator model.

The conditions the extra spectator axions ($a\geq2$) ought to satisfy in order not to impact the CMB are relatively mild for typical values of the  axion decay constant $f_a$ and energy scale $\Lambda_a$. 

\noindent Imposing (with $M_{\rm Pl}=1$)
\begin{equation}
\frac{\Lambda_a^4}{f_a}\,\sin\left(\theta_a^{\rm in}/f_a\right) \ll 3\sqrt{2\epsilon}\, H_{}^2 \;,
\label{0.3}
\end{equation}
ensures that during inflation the slope of the potential does not generate a substantial turn in field space. Moreover, when $m_a \gtrsim H_{\rm eq}$ one requires
\begin{equation}
\sum_{a\ge 2} \Lambda_a^{} f_a^{3/2}\bigl(1-\cos\left(\theta_a^{\rm in}/f_a\right)\bigr)
\;\lesssim\;
3 H_{\rm eq}^{1/2} \;,
\label{0.4}
\end{equation}
such that the combined energy fraction of axions that start oscillating before matter–radiation equality  stays small. 
Instead, when $m_a < H_{\rm eq}$, the condition
\begin{equation}
\sum_{a\ge 2}
\frac{f_a^{2}}{3}\bigl(1-\cos\left(\theta_a^{\rm in}/f_a\right)\bigr)
\;\ll\; 1 
\label{0.5}
\end{equation}
pertains to axions that start oscillating during matter domination. Finally, the inequality $\Lambda_a^2/f_a \gg H_0$ ensures axions do not behave as dark energy today and, given typical values of $f_a$ and $\Lambda_a$, is automatically satisfied.

For uniformly distributed values of $\theta^{\rm in}_{a},f_a,\Lambda_a$
and a certain number of spectator axions, say $\mathcal{O}(10)$–$\mathcal{O}(30)$, one can build a plausibility
argument for the presence of (at least) one axion in the encore regime next to a cohort of
axions with negligible impact on the CMB. Let us consider an array of $2^3$ possibilities,
\begin{equation*}
\begin{split}
&(\theta_{\rm Top},\theta_{\rm Generic}) \times 
(f_{\rm Large},f_{\rm Small})\times(\Lambda_{\rm Large},\Lambda_{\rm Small}).
\end{split}
\end{equation*}
Half of these possibilities correspond to $\Lambda_{\rm Small}$ and hence small mass and potential energy; both the slope of their potential and their energy fraction are tiny so these configurations easily satisfy the inequalities in 
\cref{0.3,0.5}. This ensures very low impact on CMB observables.

Of the remaining four classes, those with $\theta_{\rm Generic}$ will, as long as the axions remain
spectators, also have limited CMB impact. This is because, despite the fact that 
they can carry significant energy, these spectators start from a generic phase and so lose the 
logarithmic sensitivity to $\Delta \theta$ we have in the encore, which is key to the 
enhancement, as encoded in Eq.~(\ref{EnhancementintermsofN}).

The configurations $\theta_{T}f_{L}\Lambda_{L}$ and $\theta_{T}f_{S}\Lambda_{L}$ are 
the most interesting. The encore axion belongs to the first of these and it is natural 
to expect that any additional axion in this configuration may further amplify the enhancement.
The larger axion mass characterizing the $f_{S}$ configuration would 
instead lead to a faster roll on the part of the axion, one during which the field spends 
less time in the most propitious domain for the enhancement (\textit{i.e.} the one with negative 
curvature), to the detriment of its CMB impact. In fact, even a field in the $\theta_{T}f_{L}\Lambda_{L}$ 
configuration with a mass significantly larger than $H_{\rm end}$, will suffer the 
same fate for the same reason. 

The axiverse scenario may find realizations in, for example, type IIB string theory~\cite{Cicoli:2012sz,Hebecker:2018yxs,Cicoli:2021gss,Carta:2021uwv,Demirtas:2021gsq,Gendler:2023kjt,Sheridan:2024vtt,Cheng:2025ggf}, where the log uniform distribution of masses makes it possible to satisfy the conditions in \cref{0.3,0.5}.  A more thorough understanding would require the study of specific compactifications, a task we leave to future work.  

Another natural setting may be the heterotic axiverse~\cite{Agrawal:2024ejr, Reig:2025dqb, Leedom:2025mlr} where recent studies suggest that most axions acquire large masses from non-perturbative effects, thereby populating the region of parameter space directly relevant for our mechanism. In particular, axion masses from worldsheet instantons and gaugino condensation depend on the 2-cycle volumes $v$ as $m_{\theta}^2\propto {\rm exp}(-2\pi v)$. Given that weakly coupled heterotic string theory presents a bound on the overall volume of the compactification manifold ($\mathcal{V}\lesssim 25$)~\cite{Hebecker:2004ce,Cicoli:2013rwa}, individual 2-cycles cannot be parametrically large without saturating this bound, leaving little room for exponentially light axions in generic regions of moduli space. This suggests that the light-axion regime is not generic  in the heterotic axiverse.

%%%%%%%%%%%%%%%%%%%%%%%%%%%%%%%

\begin{figure*}[t!]
\centering
\begin{subfigure}{0.5\linewidth}
\centering
    \includegraphics[width=1\textwidth]{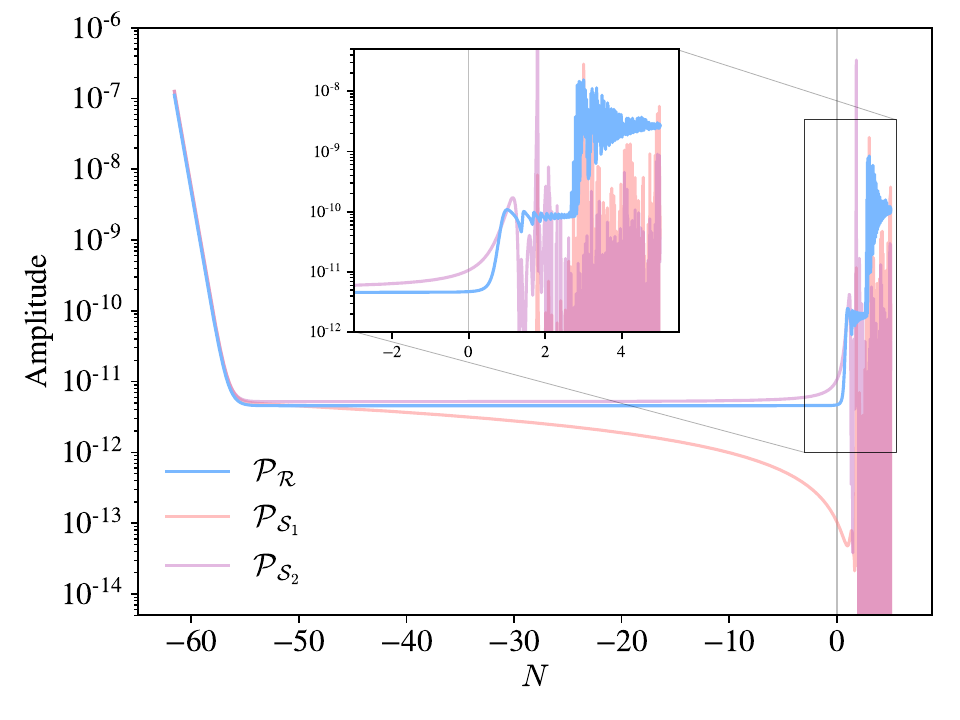}
\end{subfigure}%
\begin{subfigure}{0.5\linewidth}
\centering
    \includegraphics[width=1\textwidth]{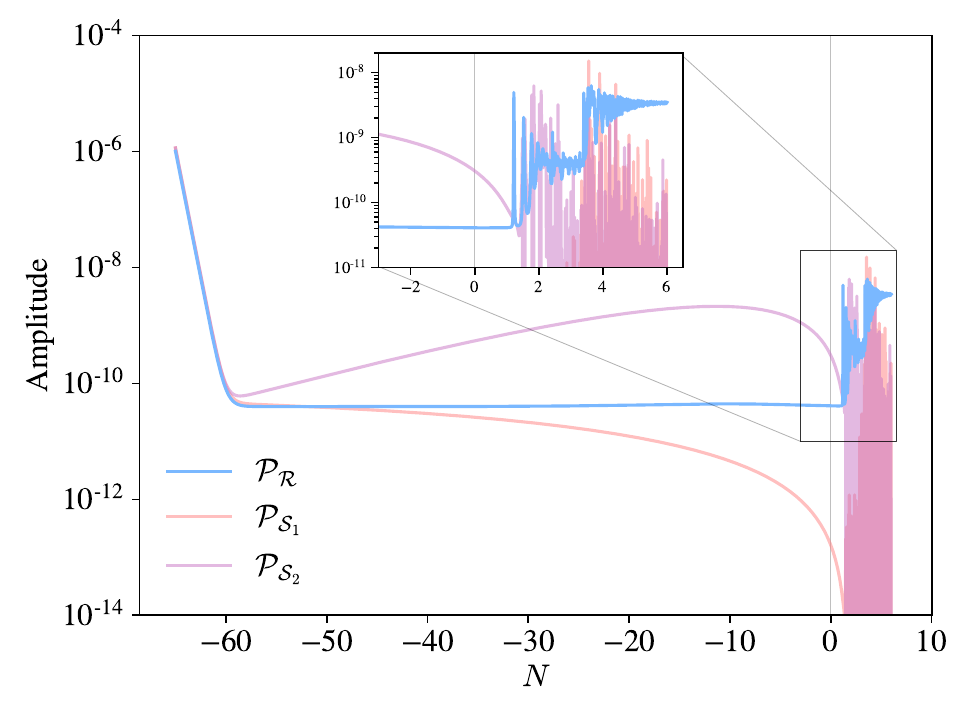}
\end{subfigure}
    \caption{Left: amplitude of the power of curvature and isocurvature fluctuations of the CMB $k$-mode for chaotic inflation with two spectator axions and $(\Lambda^4_i/m_\varphi^2, f_i^{-1},\theta_{\text{in},i}/f_i) = (10^{-4},5 \pi, 0.999\pi)$ and $(10^{-1}, \pi, 0.99\pi)$, illustrating the bursts and decays of the two isocurvature fluctuations and the sourcing of curvature modes. The overall enhancement factor approximately equals the sum of the two individual enhancement factors. Right: same plot for Starobinsky inflation, with $(\Lambda^4_i/V_0, f_i^{-1},\theta_{\text{in},i}/f_i) = (7\times10^{-3}, \pi, 0.999\pi)$ and $(10^{-4}, 3\pi, 0.9995\pi)$. Plots obtained using \texttt{PyTransport}~\cite{Mulryne:2016mzv}.
    }
    \label{fig:PowerEnhancement2}
\end{figure*}

In this work, we have focused on the simplest realization of the tachyonic encore mechanism: a gravitationally coupled two-field model with flat field-space geometry.
While chaotic and Starobinsky inflation serves as an illustrative example, our conclusions are largely model-independent. While quantitative details may differ, the ubiquity of axions in more fundamental (UV-complete) theories calls for the investigation of 
their effects on inflationary observables. From this perspective, one would not expect just an encore due to a single axion, but rather a host of such effects. As a first step in this direction, we derived the power enhancement in the presence of two similar axions alongside the inflationary sector. As befitting  gravitationally coupled sectors, the combined effect is largely additive (see \cref{fig:PowerEnhancement2}). Given the strong impact on inflationary predictions, the interplay of multiple axions with a distribution of characteristics clearly warrants further investigation. We leave this for future studies.

%%%%%%%%%%%%%%%%%%%%%%%%%%%%%%%%%%%%%%%%%%%%%%%%%%%%%%%%%%%%%%%%%%%%%%%%%%%%%%%%%%%%%%%%%%%%%%%%%%%%%%%%%%%%%%%%%%%%%%%%

% ---- Wide figure: spans both columns ----

\section{Conclusions}\label{Conclusions}
In this paper , we have presented a mechanism, based on a spectator axion initialized near the hilltop of its potential, whose post-inflationary dynamics can dramatically impact  the power spectrum, tilt, and non-Gaussianity of scalar fluctuations.~If overlaid on standard single-field slow-roll dynamics, our simple setup can reconcile well-known inflationary models with current experimental constraints. Moreover, for a large region of parameter space, it predicts order-one local non-Gaussianity, well within reach of missions currently collecting data, such as SPHEREx \cite{SPHEREx:2014bgr}.

It is worth emphasizing that, in contradistinction to standard mechanisms such as the curvaton~\cite{Lyth:2001nq,Enqvist:2001zp,Moroi:2001ct,Bartolo:2002vf,Bartolo:2001rt,Fonseca:2012cj,Byrnes:2025kit} or modulated reheating~\cite{Dvali:2003em,Dvali:2003ar,Kofman:2003nx,Benaco:2025tlp}, the shift 
in $(n_s,r)$ and the generation of observable $f^{\rm loc}_{\rm NL}$ in the encore 
scenario do not rely on an additional conversion mechanism, such as a decay channel or a modulated reheating surface. Both the curvature enhancement and its nonlinear signature stem from the same post-inflationary 
axion dynamics.

\begin{acknowledgments}
We are delighted to thank Ana Ach\'{u}carro, Renata Kallosh, Andrei Linde, Marieke Postma, David Wands and Alexander Westphal for illuminating conversations.~RGQ thanks CONAHCyT (SECIHTI) and RUG for funding and support, and also acknowledges support from JST EXPERT-J, Japan Grant No. JPMJEX2508, as well as support for the Kavli IPMU from the World Premier International Research Center Initiative (WPI Initiative), MEXT, Japan.~MP's work is supported by the NWA through the ORC programme
Emergence at All Scales.~MF acknowledges CSIC grant 20235AT005. MF's work is  supported by the
Agencia Estatal de Investigaci\'{o}n through the Grant IFT Centro de Excelencia Severo Ochoa No CEX2020-001007-S, funded by 
MCIN/AEI/10.13039/501100011033.
\end{acknowledgments}

\appendix
\section{Reheating}\label{Reheating}

A possible concern regarding the encore mechanism is that our analysis neglects the effects of reheating, despite relying on post-inflationary dynamics. Here we show that this assumption is justified within standard perturbative reheating scenarios. 

We model a given perturbative reheating process through the inclusion of a decay term in the inflaton equation of motion with a flat internal metric
\begin{equation}
    \ddot{\varphi} + (3H+\Gamma)\dot{\varphi} + V_{,\varphi}=0\,,
\end{equation}
where $\Gamma$ is the inflaton decay rate. Reheating becomes dynamically relevant the moment $\Gamma \sim H$, while for $\Gamma \ll H$ its effect is subdominant and the post-inflationary dynamics proceed as in the absence of reheating, exactly as in our analysis. 

After inflation, the inflaton undergoes an oscillatory phase characterized by an effective equation of state $w_{\rm eff}$. The Hubble parameter is then given by 
\begin{equation}
    H(N_{\rm post})=H_{\rm end}\ \text{exp}\left[-\frac{3}{2}(1+w_{\rm eff})N_{\rm post}\right]\,,
\end{equation}
with $N_{\rm post}$ the number of $e$-folds elapsed since the end of inflation. Requiring reheating to remain negligible during this period implies $\Gamma\ll H(N_{\rm post})$.
Using the standard perturbative reheating relation \cite{McDonald:1999hd,Kolb:2003ke}
\begin{equation}
    T_{\rm re}\simeq
    \left(\frac{90}{\pi^2 g_*}\right)^{1/4}\sqrt{\Gamma }\,,
\end{equation}
where we set $M_{\rm Pl}=1$ and where $g_{*}$ is the number of relativistic species at a given temperature. We thus obtain the temperature at which reheating becomes dynamically relevant
\begin{equation}
\begin{split}
    T^{\rm dyn}_{\rm re}\simeq{}&
    \left(\frac{90}{\pi^2 g_*}\right)^{1/4}
    \sqrt{H_{\rm end}}\exp\left[-\frac34(1+w_{\rm eff})N_{\rm post}\right]\,.
\end{split}
\end{equation}
In the present setup, the relevant scale is set by the spectator axion dynamics with $|m_\theta|\sim H_{\rm end}$ so the same result can equivalently be written in terms of the axion mass. 
For a matter-like post-inflationary phase (such as the one in the two examples in the main text) $w_{\rm eff}=0$, assuming that all the Standard Model species are relativistic at the end of inflation, $g_*\sim100$, and considering a typical inflationary scale, $H_{\rm end}\sim10^{13}\,\mathrm{GeV}$, we find 
\begin{equation}
    T^{\rm dyn}_{\rm re}\simeq3\times10^{15}e^{-3N_{\rm post}/4}\,\mathrm{GeV}\,.
\end{equation}
In order to safely neglect any reheating dynamics, we thus limit the possible reheating scenarios to those that give rise to a temperature satisfying $T_{\rm re}\ll T_{\rm re}^{\rm dyn}$. This condition still leaves a large viable window for reheating processes. Indeed, successful Big Bang nucleosynthesis only requires $T_{\rm re}\gtrsim \mathcal{O}(10)\,\text{MeV}$, while for $N_{\rm post}=10$ one finds $T_{\rm re}^{\rm dyn}\sim10^{12}\,\text{GeV}$. Therefore, reheating temperatures several orders of magnitude below $T_{\rm re}^{\rm dyn}$ remain comfortably compatible with the lower bound from standard cosmology. Consequently, the axion encore dynamics can consistently unfold entirely during a transient post-inflationary epoch prior to the onset of efficient reheating.

\bibliographystyle{JHEP}
\bibliography{biblio}
\end{document}